# The Journal Coverage of Web of Science and Scopus: a Comparative Analysis

Philippe Mongeon and Adèle Paul-Hus
philippe.mongeon@umontreal.ca; adele.paul-hus@umontreal.ca
Université de Montréal, École de bibliothéconomie et des sciences de l'information,
C.P. 6128, Succ. Centre-Ville, H3C 3J7 Montréal, Qc, Canada

## Abstract

Bibliometric methods are used in multiple fields for a variety of purposes, namely for research evaluation. Most bibliometric analyses have in common their data sources: Thomson Reuters' Web of Science (WoS) and Elsevier's Scopus. This research compares the journal coverage of both databases in terms of fields, countries and languages, using Ulrich's extensive periodical directory as a base for comparison. Results indicate that the use of either WoS or Scopus for research evaluation may introduce biases that favor Natural Sciences and Engineering as well as Biomedical Research to the detriment of Social Sciences and Arts and Humanities. Similarly, English-language journals are overrepresented to the detriment of other languages. While both databases share these biases, their coverage differs substantially. As a consequence, the results of bibliometric analyses may vary depending on the database used.

## Introduction

Bibliometric and scientometric methods have multiple and varied application realms, that goes from information science, sociology and history of science to research evaluation and scientific policy (Gingras, 2014). Large scale bibliometric research was made possible by the creation and development of the Science Citation Index (SCI) in 1963, which is now part of Web of Science (WoS) alongside two other indexes: the Social Science Citation Index (SSCI) and the Arts and Humanities Citation Index (A&HCI) (Wouters, 2006). The important feature of these databases is that they include all article types and index all authors, institutional addresses and bibliographic references for each article. WoS had been the sole tool for citations analysis until the creation of Scopus and Google Scholar in 2004. However, the low data quality found in Google Scholar raises questions about its suitability for research evaluation. Thus, WoS and Scopus remain today the main sources for citation data. Moreover, the interdisciplinary coverage of these databases represents a significant strength for the study and comparison of different scientific fields (Archambault et al., 2006).

**Limits of databases for the evaluation of research in social sciences and humanities**

The validity of bibliometric analyses for research evaluation lies in large part on the databases' representativeness of the scientific activity studied. One of the main issues with citation indexes like WoS and Scopus is that their coverage mainly focuses on journals and less on other means of scientific knowledge diffusion (e.g., books, proceedings and reports). This can be problematic since scientific communication practices are largely influenced by the research field "epistemic cultures" (Knorr-Cetina, 1991). Indeed, while the article is the dominant mean of results dissemination in Natural Science, Engineering and Biomedical Research, it is not the case in many Social Sciences disciplines and in an even more pronounced way in Arts and Humanities where publishing books is more frequent and more important for researchers' career than publishing articles (Glänzel and Schoepflin, 1999; Larivière et al., 2006). As a consequence, the portrait of scientific output and impact—at any level of aggregation (individual, institutional or national) – that WoS and Scopus provide cannot be as accurate for Social Sciences and Arts and Humanities as it may be for Natural Sciences, Engineering and Biomedical Research (Hicks and Wang, 2011; Nederhof, 2006). It should be noted that both WoS and Scopus have tried addressing this issue; Thomson Reuters, the corporation owning WoS, by creating its book citation index, and Elsevier, owning Scopus, by recently adding books to its database coverage.

A second important issue is the language coverage of citation databases. More than a decade ago, van Leeuwen et al. (2001) were advising for caution in interpreting bibliometric data in comparative evaluation of national research systems as a consequence of the language biases of the WoS Science Citation Index coverage. Analyzing data from 2004, Archambault et al. (2006) also observed an important English-language journals overrepresentation in the WoS coverage compared to Ulrich's database, which is considered the most comprehensive worldwide list of periodicals. They concluded that "Thomson Scientific databases cannot be used in isolation to benchmark the output of countries in the [Social Sciences and Humanities]" (p. 329).

**Literature review**

Archambault et al. (2009) have shown a high correlation between the number of papers and the number of citations received by country calculated with Scopus and with the WoS, and thus concluded that both databases are suitable tool for scientometrics analyses. Gavel and Iselid (2008) analyzed the journal coverage overlap between Scopus and WoS, based on 2006 data and showed that, at the time, 54% of active titles in Scopus were also in WoS and that 84% of active titles in WoS were also indexed in Scopus.

Several studies have measured the overlap between databases and the impact of using different data sources for specific research fields on bibliometric indicators. For Earth Sciences, Mikki (2009) compared Google Scholar coverage to WoS: 85% of the literature indexed by WoS in that field was recalled by Google Scholar. Barnett and Lascar (2012) found that Scopus had more unique journal titles than WoS in the field of Earth and Atmospheric Sciences. However, in both databases, unique titles had low Journal Rank Indicators (Scopus) and Impact Factors (WoS), thus indicating a minor role of the data source within that specific field. De Groote and Raszewski (2012) performed a similar analysis for the field of Nursing. Comparing the coverage of WoS, Scopus and Google Scholar to calculate h-index of a sample of authors, they conclude that more than one tool must be used in order to provide a thorough assessment of a researcher's impact. In the field of Business and Management, Mingers and Lipitakis (2010), and Clermont and Dyckhoff (2012) showed that Google Scholar mainly indexes international, English-language journals and while it includes unreliable data, it has a better coverage compared to Scopus and WoS.

Meho and Yang (2007) compared citation data and ranking of scholars in the field of Library and Information Science using WoS, Scopus and Google Scholar. Their results show that Google Scholar had, at the time, the most extensive coverage of conference proceedings and non-English language journals. They conclude that the use of Scopus and Google Scholar in addition to WoS contribute to a more accurate assessment of authors' impact. Abrizah et al. (2012) also analyzed the journal coverage of Library and Information Science in WoS and Scopus and found a total of 45 titles covered in both databases with normalized impact factors being higher for titles covered in Scopus. Furthermore, Scopus covered more unique titles (n=72) than did WoS (n=23). For Computing Sciences, Franceschet (2009) concluded that Google Scholar compiles significantly higher indicators' scores than WoS. However, rankings based on citations data from both databases are not significantly different. López-Illescas et al. (2008) compared oncological journals coverage in WoS and Scopus and found that Scopus covered a larger number of titles. Nevertheless, 94% of Scopus highest impact factor journals were indexed in WoS. Comparing WoS and Google Scholar citations in four different scientific fields (Biology, Chemistry, Physics and Computing), Kousha and Thelwall (2007) found that the majority of Google Scholar unique citations (70%) were from full-text sources. Moreover, types of citing documents significantly differed between disciplines thus suggesting that a large range of academic non-journal publications are not indexed in WoS but are accessible through Google Scholar. Other studies have looked at how well these databases cover the scientific output of specific countries or regions, such as Spain (Psychology research in Spain) (Osca-Lluch et al., 2013) and Latin America and the Caribbean (Santa and Herrero-Solana, 2010).

Since the journal coverage of the WoS and Scopus are not static and evolved through time, we propose in this paper to revisit the question and to compare the coverage of both WoS and Scopus in terms of fields, countries and languages, thus examining if the previously found biases in WoS databases still exist and if similar biases can be found in Scopus. This study aims at providing a complete and up-to-date portrait of the journal coverage of WoS and Scopus by comparing their coverage with one another and with Ulrich's extensive periodical directory.

# Methodology

## Journal lists

We searched the online version of the Ulrich's periodical database for all journals classified as "Academic/Scholarly". 162, 955 records corresponded to this criterion and were downloaded manually[1]. This was a lengthy process as a maximum of 500 records can be downloaded at once. Ulrich's database often contains multiple entries for a single journal. This is the case, for example, when a journal is published in more than one format (eg. online and print), or in more than one language. After eliminating those duplicate entries, 70,644 unique journals remained.

The Thomson Reuters' master journal list was downloaded from the Thomson Reuters Website[2]. However, Thomson Reuters doesn't provide its journal list in a spreadsheet (e.g. Excel), we thus downloaded the source code from the website and used a XSL style sheet to convert the data into a table. The list combines journals indexed in the Science Citation Index Expanded (SCI-Expanded), the Social Science Citation Index (SSCI) and the Arts and Humanities Citation Index (A&HCI). The master journal list totalled 16,957 entries[3]. We downloaded the list of journals indexed in Scopus from the Elsevier website[4], which is provided directly in a spreadsheet format. The list contained 34,274 titles and, according to the website, it was last updated in May 2014.

## Matching

Journals from the WoS and the Scopus lists were matched to the Ulrich's list in two steps. In the first step, the journals were matched by means of their ISSN. In the second step, the remaining journals were matched by means of their title, and these matches were manually verified to eliminate any false positives. Using this procedure, we were able to match 14,637 and 23,189 journals from the WoS and Scopus lists, respectively, with the Ulrich's list.

We decided to limit our analysis to active journals and, thus, limited our matches to titles which were classified as "serial type = journal" and "status = active" in Ulrich. 296 matched journals from WoS had a status other than active in Ulrich. We manually verified the status of these journals and found that 21 were in fact still active. Thus, we considered these 21 journals as active and updated the Ulrich's list accordingly. Similarly, 363 matched journals from Scopus had an inactive status in Ulrich. We manually verified the 363 journals and found a total of 197 active journals and thus updated the Ulrich's list accordingly. Following these operations, our final sample comprised 13,605 and 20,346 matched journals from WoS and Scopus, respectively, on a total of 63,013 active journals in Ulrich.

---

[1] http://ulrichsweb.serialssolutions.com/. Data downloaded bewteen June 8th and June 12th 2014.
[2] http://ip-science.thomsonreuters.com/mjl/. Data Downloaded on June 25th 2014.
[3] The list included inactive journal titles, namely from the Zoological Record and BIOSOS Previews collections.
[4] http://www.elsevier.com/online-tools/scopus/content-overview. Downloaded on June 25th 2014.

## Journals classification

### By field

A broad discipline classification of journal titles was done by assigning every Ulrich's subject to one of the four broad field of the National Science Foundation (NSF) classification (NSF, 2006): Natural Sciences and Engineering, Biomedical Research (which includes the biomedical research and clinical medicine NSF categories, but not health, which is part of social sciences), Social Sciences, and Arts and Humanities.

### By country

To assign a country to each journal, we used the publisher's country provided by Ulrich, which is the only geographical information available in that database. Most journals had only one publisher's country, but some had two or more (e.g. the print version is published in a country and the online version in another). In these cases, all countries were assigned to the journal, using the full counting method, which means that a journal published in Canada and the United States will count as one journal for Canada and one journal for the United States, as opposed to fractional counting in which case this journal would be counted as 0.5 journal for each country.

### By language

The assignment of one or more language to each journal was done by using the data provided in the journal's record in Ulrich, using the full counting method. Some journals had abstracts, comments or notes in an additional language than the text of the articles. In those cases, we only included the language of the text. In other words, we counted only the language in which the full texts of the journals are written.

## Data analysis

### Databases coverage

The relative coverage of WoS and Scopus was calculated by dividing the number of journals of each database with the number of journals in Ulrich. This was done for each of the four broad fields of research.

### Relative distribution of journals by field, country and language

In order to assess whether or not there is an overrepresentation or an underrepresentation of a field within a database, we compared distributions of journals in WoS and Scopus with the distribution of journals in Ulrich. An overrepresentation or underrepresentation might indicate that cross-field comparison of scientific output and impact using these databases may favor some fields to the detriment of others. For each field, the relative distribution of journals was calculated by dividing the proportion of journals of each field in WoS and Scopus by the proportion of those fields in Ulrich. This was also done for countries and languages within each field.

### Coverage overlap of Web of Science and Scopus

In order to assess the extent by which the WoS and Scopus journal coverage overlap, we classified each journals in the three following categories: WoS only (journals that are indexed in WoS but not in Scopus),

Overlap (journals that are indexed in both WoS and Scopus), and Scopus only (journals that are indexed in Scopus but not in WoS).

## Results

### Coverage by field

Figure 1 shows the proportion of Ulrich's journals covered by WoS and Scopus within each field. The largest coverage difference appears in Biomedical Research (BM), with Scopus covering almost half of all Ulrich's journals in this field while WoS covers only about 28%. It is in Natural Sciences and Engineering (NSE) the coverage is the most similar between the two databases, Scopus covering 38% of journals and WoS, 33%. NSE is also the field where WoS has the highest coverage. We also notice that Social Sciences (SS) and Arts and Humanities (AH) are not covered as well as the two other fields. Indeed, Scopus covers less than 25% of journals in both fields, while WoS covers less than 15%.

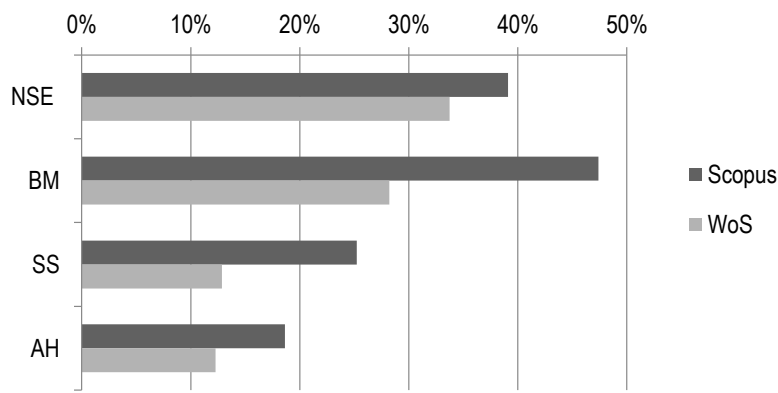

**Fig 1.** Proportion of Ulrich academic journals indexed in Web of Science and Scopus

Table 1 shows the distribution of journals by field within each of the database. We see that NSE is well overrepresented in WoS ($\cong$ 43% of WoS journals compared to $\cong$ 28 % of Ulrich journals) and also but in a lesser extent in Scopus ($\cong$ 33%). It is also the case for BM, which is overrepresented both in WoS ($\cong$ 27%) and Scopus ($\cong$ 31%), if we compare to Ulrich distribution ($\cong$ 21%). The Scopus coverage shows a stronger overrepresentation of BM journals and WoS, a stronger overrepresentation of NSE journals. Social Sciences (SS), which have the biggest share of journals in Ulrich ($\cong$ 36%), are underrepresented in WoS ($\cong$ 21%) and in Scopus ($\cong$ 28%). Finally, AH ($\cong$15% in Ulrich) are also underrepresented in both WoS and Scopus ($\cong$ 9%).

**Table 1.** Relative distribution of journals by discipline in Ulrich, Web of Science and Scopus

| Field | Ulrich | | WoS | | | Scopus | | |
|---|---|---|---|---|---|---|---|---|
| | N | % | N | % | Difference | N | % | Difference |
| NSE | 17,213 | 27,5% | 5,810 | 42,7% | 55,1% | 6,730 | 32,9% | 19,5% |
| BM | 13,232 | 21,2% | 3,732 | 27,4% | 29,6% | 6,271 | 30,6% | 44,8% |
| SS | 22,519 | 36,0% | 2,893 | 21,3% | -41,0% | 5,682 | 27,8% | -22,9% |
| AH | 9,559 | 15,3% | 1,172 | 8,6% | -43,7% | 1,781 | 8,7% | -43,1% |
| Total | 62,523 | 100,0% | 13,607 | 100,0% | 0 | 20,464 | 100,0% | 0 |

Figure 2 shows the overlap in the journal coverage of both databases. Overall, except for NSE, Scopus includes most of the journals indexed in WoS. Furthermore, Scopus has a larger number of exclusive journals than WoS in all fields, which can be explained by the fact that Scopus covers a lot more journals than WoS (Table 1). While this is true for all fields, NSE is the field where WoS has the highest number of exclusive journals, compared to other fields.

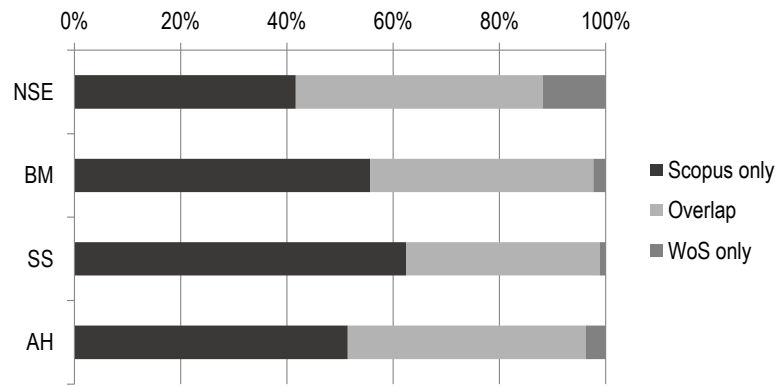

**Fig 2.** Coverage overlap of Web of Science and Scopus, by discipline

## Coverage by country of publisher

As mentioned in the methods section, the journal publisher's country was used to assign a country to each journal, a choice that introduces some limitations to our analysis. To illustrate theses limits Table 2 presents the number of articles published by author's country and the number of journals by publisher's country. It shows that the most important producers of scientific papers are not necessarily the most important publishers. For example, the Dutch authors published about 2% of all paper in 2013 but about three times more journals were published in the Netherlands.

**Table 2.** Number of articles and journals indexed in Web of Science and Scopus for the top 15 countries in terms of articles indexed

| Country | Articles | | | | | | Jounals | | | | | |
|---|---|---|---|---|---|---|---|---|---|---|---|---|
| | WoS | | | Scopus | | | Wos | | | Scopus | | |
| | N | % | Rank | N | % | Rank | N | % | Rank | N | % | Rank |
| USA | 352,477 | 23.2 | 1 | 451,292 | 22.5 | 1 | 4,176 | 30.7 | 1 | 5,858 | 28.4 | 1 |
| China | 207,979 | 13.7 | 2 | 322,041 | 16.0 | 2 | 269 | 2.0 | 6 | 489 | 2.4 | 6 |
| UK | 108,455 | 7.1 | 3 | 132,615 | 6.6 | 3 | 3,293 | 24.2 | 2 | 4,738 | 23.0 | 2 |
| Germany | 95,267 | 6.3 | 4 | 117,184 | 5.8 | 4 | 959 | 7.0 | 3 | 1,241 | 6.0 | 4 |
| Japan | 73,878 | 4.9 | 5 | 94,015 | 4.7 | 5 | 303 | 2.2 | 5 | 454 | 2.2 | 6 |
| France | 66,222 | 4.4 | 6 | 83,692 | 4.2 | 7 | 250 | 1.8 | 7 | 510 | 2.5 | 5 |
| Canada | 58,378 | 3.8 | 7 | 72,422 | 3.6 | 9 | | N/A* | | | | N/A* |
| Italy | 58,119 | 3.8 | 8 | 73,047 | 3.6 | 8 | 227 | 1.7 | 9 | 366 | 1.9 | 9 |
| Spain | 51,829 | 3.4 | 9 | 65,571 | 3.3 | 10 | 170 | 1.2 | 15 | 406 | 2.0 | 8 |
| Australia | 49,462 | 3.3 | 10 | 62,910 | 3.1 | 11 | 215 | 1.6 | 10 | 315 | 1.5 | 12 |
| India | 48,591 | 3.2 | 11 | 85,100 | 4.2 | 6 | 200 | 1.5 | 11 | 436 | 2.1 | 7 |
| S. Korea | 47,949 | 3.2 | 12 | 58,425 | 2.9 | 12 | | N/A* | | | | N/A* |
| Brazil | 35,684 | 2.3 | 13 | 50,710 | 2.5 | 13 | 176 | 1.3 | 14 | 267 | 1.3 | 15 |
| Netherlands | 35,153 | 2.3 | 14 | 42,296 | 2.1 | 14 | 927 | 6.8 | 4 | 1,498 | 3 | 3 |
| Russia | 27,313 | 1.8 | 15 | 38,045 | 1.9 | 15 | 193 | 1.4 | 12 | 314 | 1.5 | 13 |

*Canada and South Korea do not appear in the top 15 journal publishing countries. Switzerland (ranked 8[th] in WoS and 11[th] in Scopus) and Poland (ranked 13[th] in WoS and 14[th] in Scopus).

On the other hand, China is the second biggest producer of articles (≅ 15%) but published only 2% of all journals. Thus, major actors in terms of scientific production are not the same as major actors in terms of academic publishing. Furthermore, while the article and journal relative coverage ranking are overall similar in WoS and Scopus, Table 2 shows that Spanish and Indian journals, for example, have a much larger relative coverage in Scopus than in WoS.

Figure 3 shows the relative distribution (left scale) and the absolute number (right scale) of journals covered in the WoS and Scopus compared to Ulrich for the top 15 publishing countries. The zero on the relative distribution scale represents the coverage in Ulrich. Countries above zero are overrepresented in WoS or Scopus compared to Ulrich, while those under zero are underrepresented. In all four fields, the same countries (i.e. the Netherlands, the United Kingdom, and the United States) are overrepresented in every field. This is not surprising considering that some of the major academic publishing companies are located in these countries (e.g. Elsevier in the Netherlands and Sage and Routledge in the UK). In turn, most of the other countries are underrepresented.

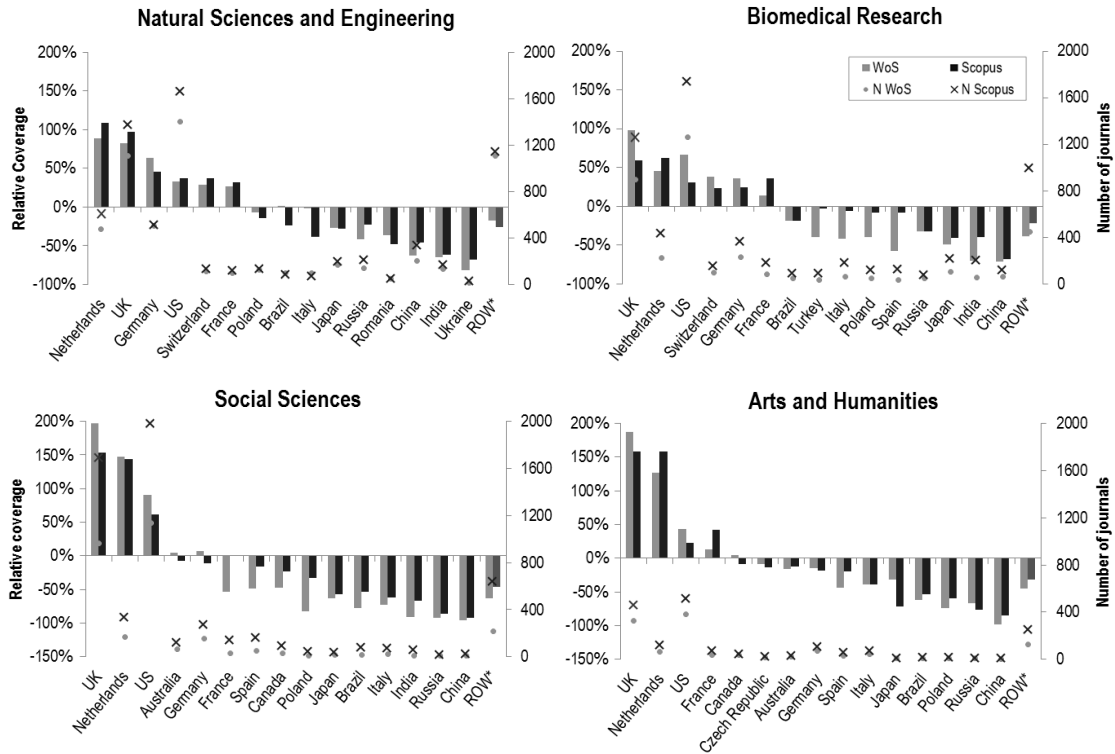

**Fig. 3.** Relative distribution and number of journals covered in Web of Science and Scopus by publisher's country

Even though WoS and Scopus have similar relative coverage in terms of publishing countries, they do not necessarily index the same journals. Figure 4 shows the overlap in journal coverage between WoS and Scopus in terms of publishing countries for each of the four fields. As mentioned previously, because of the larger number of journals in Scopus (Table 1), Scopus has a larger proportion of exclusive journals (i.e. journals that are not indexed in WoS), and this is the case in all fields. In fact, especially in BM, SS and AH, most of WoS journals are also covered by Scopus. In NSE, 56% of journals published in China and Russia are only indexed in Scopus. Brazil shows a more balanced situation with 31% of journals exclusive to WoS, 32% exclusive to Scopus and 37% of overlap. On average for the top 15 publishing countries in NSE, 13% of journals are exclusive to WoS (ranging from 0% to 31%) and Scopus has 41% of exclusive journals (ranging from 22% to 56%), with an overlap ranging from 22% to 78%.

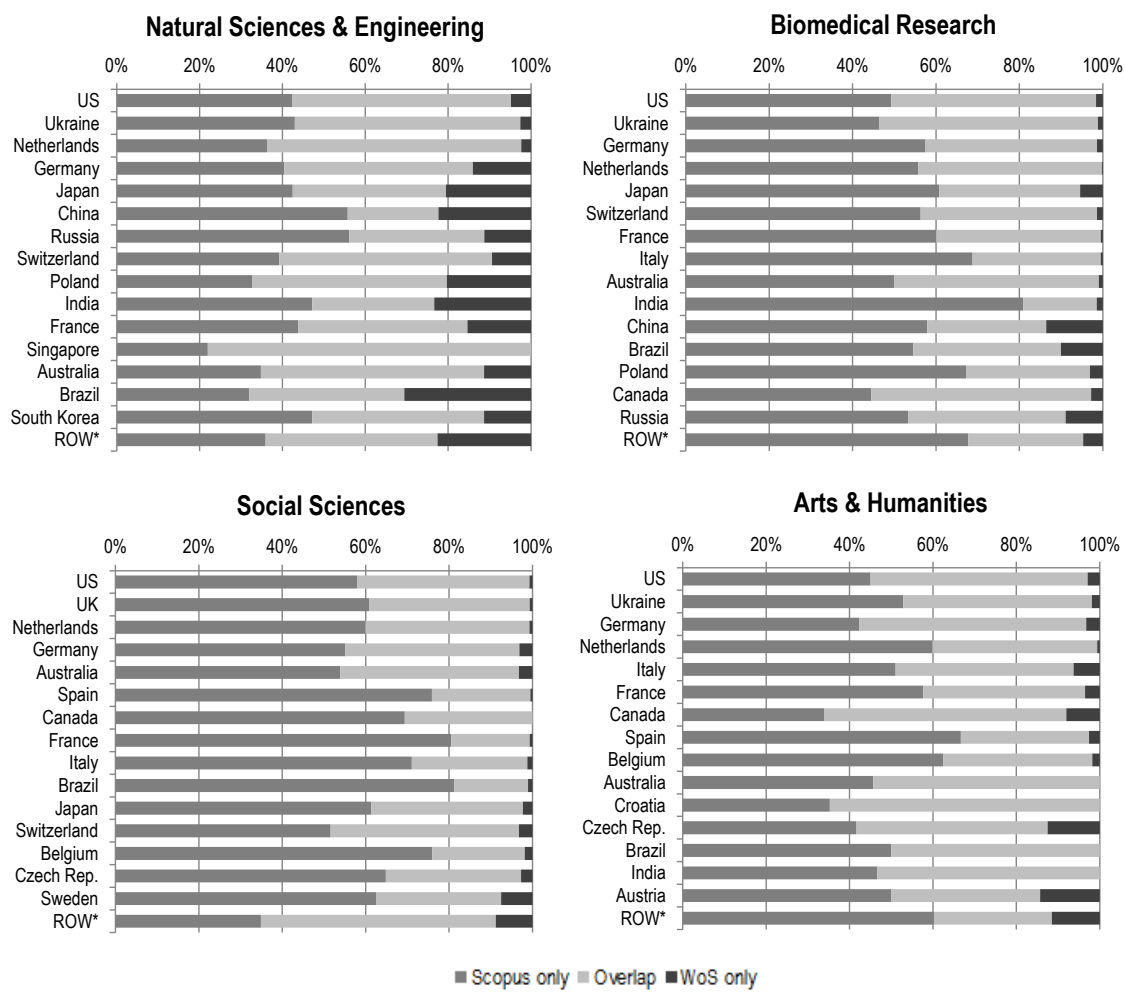

*: aggregated data for the rest of the world (ROW)

**Fig. 4.** Journal coverage overlap in Web of Science and Scopus by publishing country

In BM, more than 60% of journals published in Japan, Italy and Poland, and 81% of journals published in India are exclusive to Scopus. On average for the top 15 countries, 58% of journals are exclusive to Scopus (ranging from 53% to 81%) while an average of 4% of journals are only indexed in WoS (ranging from 0% to 14%). The average overlap is 39%, ranging from 18% to 53% within the top 15 countries. Journals of SS have an even skewer distribution for the top 15 countries. On average, 64% of journals are exclusive to Scopus (ranging from 52% to 81%), with 81% of journals published in France and in Brazil only indexed in Scopus. The overlap averages 34% (ranging from 18% to 43%), while an average of 2% of journals are exclusive to WoS (ranging from 0% to 8%). In AH, 67% of journals published in Spain are only indexed in Scopus. On average for the top 15 countries, 49% of journals are exclusive to Scopus (ranging from 34% to 67%), 47% of journals are indexed in both databases (with an overlap ranging from 31% to 65%) while only 4% of journals are exclusive to WoS (ranging from 0% to 14%).

## Coverage by language

Figure 5 presents the 15 most frequent languages in Ulrich, and shows that for all fields in Scopus, the majority of languages are underrepresented, and WoS shows a similar trend with the exception of NSE, where nine languages are overrepresented. As we would expect, since English has a dominant position in sciences, English is overrepresented in the four fields. It is the only language that is constantly and strongly overrepresented in the two databases and in all fields. Interestingly, Dutch is the most overrepresented language in NSE in both databases. It should be pointed out, however, that we count less than 50 Dutch journals in NSE.

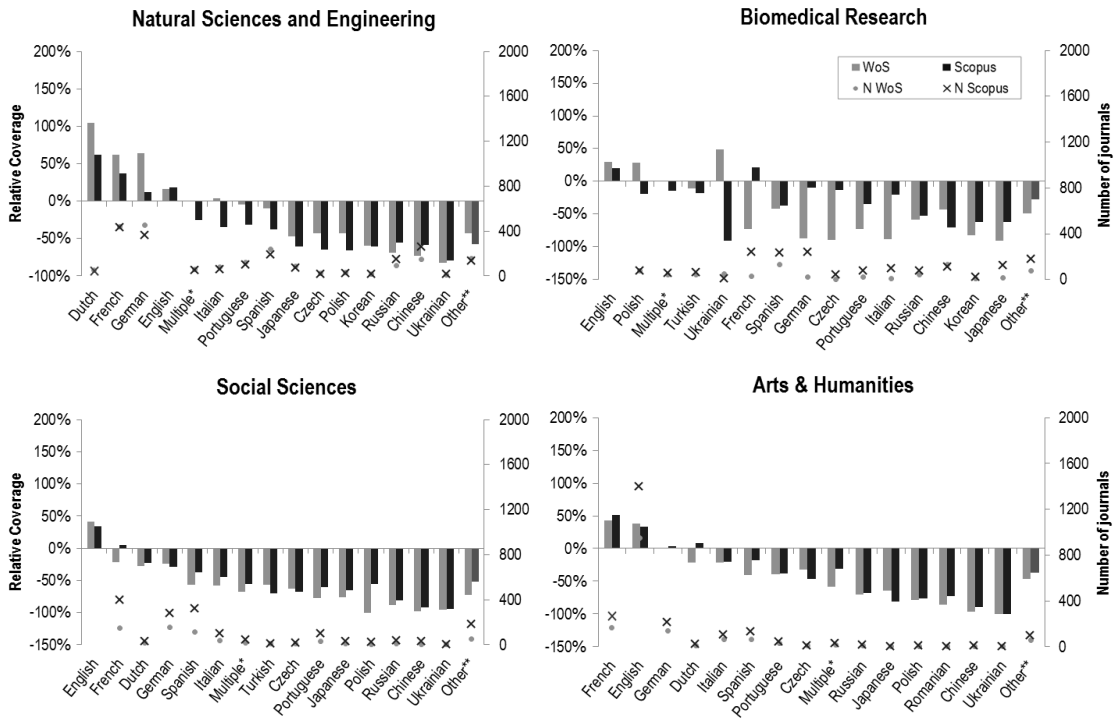

**Fig. 5.** Relative distribution and number of journal languages covered in Web of Science and Scopus

WoS and Scopus show similar trends (i.e. the same languages are over or underrepresented) in terms of language coverage for NSE, SS and AH. A few notable exceptions are French in SS and Dutch in AH, which are both overrepresented in Scopus and underrepresented in WoS. The field of BM presents a different profile with highly discrepant results. For example, Ukrainian is well overrepresented in WoS and well underrepresented in Scopus, while French, on the opposite, is well overrepresented in WoS and underrepresented in Scopus. One should however note, that there is only about 100 biomedical journals in Ukrainian (less than 1% of all journals analyzed).

Figure 6 shows the coverage overlap between WoS and Scopus in terms of journal language. As it was the case for countries (Figure 4), the higher number of journals indexed in Scopus explains that this database has a more important share of exclusive journals and covers most of the journals indexed in the WoS.

Again, NSE is the exception with an important proportion of non-English journals only covered in WoS. However, more than 55% of NSE journals in Russian and in Chinese are only indexed in Scopus. On average, for the top 15 languages, 34% of journals are exclusive to Scopus (ranging from 7% to 59%), 39% are indexed in both Scopus and WoS (ranging from 14% to 81%) and 28% are exclusive to WoS (ranging from 8% to 41%).

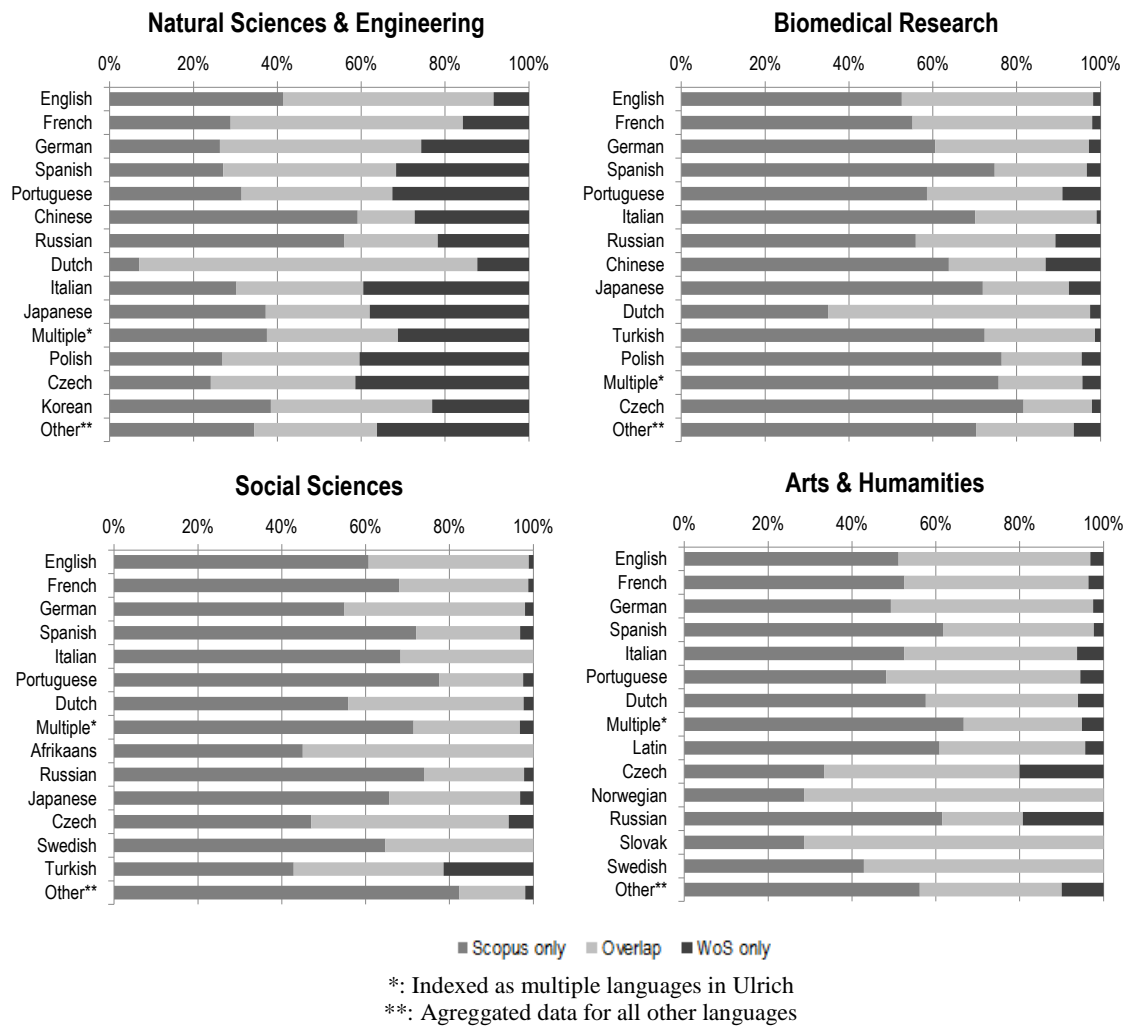

*: Indexed as multiple languages in Ulrich
**: Agreggated data for all other languages

**Fig. 6.** Journal coverage overlap in Web of Science and Scopus by language

The BM journals show a highly skewed distribution in terms of language overlap. On average for the 15 most frequent languages, 65% of journals are exclusive to Scopus (ranging from 35% to 82%) while only 5% of journals are exclusive to WoS (ranging from 1% to 13%), with an overlap ranging from 16% to 63%. The SS field shows a similar situation with an average of 62% of journals exclusive to Scopus (ranging from 43% to 78%), 3% exclusive to WoS (ranging from 0% to 21%) and an overlap of 35% (ranging from 20% to 55%), for the top 15 languages. In AH, on average, 50% of journals are exclusive to Scopus, (ranging from 29% to 67%), 45% are indexed in both databases (ranging from 19% to 71%) while 6% are only exclusive to WoS (ranging from 0% to 20%).

## Discussion

The results of this study show that, as Archambault et al. (2006) had found, there is still an overrepresentation of certain countries and languages to the detriment of others in the WoS journal coverage. Similar biases were found in the coverage of Scopus, despite its much larger journal coverage. Overall, journals published in countries like the United States, the United Kingdom, the Netherlands, France, Germany and Switzerland represent a larger proportion of journals indexed in WoS and Scopus than they do in Ulrich. A potential explanation for this finding could be that these countries have a longer history in academic publishing and have more research resources (e.g., funding, infrastructures, and institutions) than smaller or developing countries. Also some of the big commercial academic publishers (e.g. Elsevier, Springer, Taylor & Francis, Wiley-Blackwell, Sage Publications) are located in these countries, and as Larivière et al. (in press) have shown, more than 50% of all articles indexed in WoS are published in journals own by those five commercial publishers.

The results regarding coverage overlap provide an estimation of the tool-dependency level when comparative analyses are performed with those data sources. The WoS and Scopus journal coverage differs the most in Natural Science and Engineering and in Arts and Humanities (lowest overlap average, and highest proportion of WoS exclusive journals). For example, analyses of Chinese journals in the field of NSE may vary greatly depending on the data source, as only 21% of journals are indexed by both WoS and Scopus, while similar analyses of Singaporean journals may lead to more coherent results as 78% of journals covered by both databases.

The main limitations of this study come from the definition of the country and language(s) of a journal. Defining the geographical origin of a journal was not a straightforward process. What criteria should be used to determine the country of a journal? Should it be determined by the country of its publisher or of its editor? Perhaps it should be determined by the location of all the authors signing an article? Given data availability, publisher's country was used even though the information it conveys could be argued to be rather financial. In terms of languages definition, another problem emerged as some journals had quite a long list of languages in Ulrich's language field, while others only had the mention "multiple languages". It is unclear how many languages a journal must be published in to fall in this category, or if that distinction is due to inconsistencies in the data. For example, there were cases where a single journal had: English, French, German, Spanish and Dutch listed as language. This raises questions about what actually differentiate journals with long lists of languages and those with the "multiple languages" mention. While this is an obvious limit of our dataset, its' potential impact on the results is minimal since these "multiple languages" cases only represent 1.4% of all the journals indexed in Ulrich.

## Further research

This study has given a portrait of WoS and Scopus journal coverage based on field classification, publisher's country and language of journals, information provided in the Ulrich's periodical database. Further research could look at the language at the article-level of a journal to better grasp the proportional language coverage of a journal. In Social Sciences and Arts and Humanities it would be relevant to investigate if national journals which focus on local matters are well represented in WoS and Scopus. Or do these databases have a clear international focus? Further research could also investigate the notion of national versus international identity of journals, in terms of subjects, editorial board, and the nationality of authors who published in those journals, since many journals, like Science, aim to be international, while 90% of the articles they publish come from American authors.

## Conclusion

Our analysis shows that the journal coverage of WoS in Social Sciences and Arts and Humanities is still quite low and that these disciplines are underrepresented as compared to their share in Ulrich. Also, the strong English-language overrepresentation in WoS found by Archambault et al. (2006) proved to be persistent as it is confirmed by our data, some ten years later. We also found that despite Scopus's larger journal coverage in all fields, the database shows similar biases than those found in WoS. Consequently, using WoS and Scopus for research evaluation introduces biases that favor Natural Sciences and Engineering as well as Biomedical Research to the detriment of Social Sciences and Arts and Humanities. Similarly, English-language journals are favored to the detriment of other languages. These important limits should be taken into account when assessing scientific activities. They also raise the question as to whether there are better tools that could be used for research evaluation. For example, Google Scholar provides free access to scholarly documents of all types, language and for all fields. It is widely used for information retrieval, but its suitability for research evaluation and other bibliometric analyses has been highly questioned because of the sporadic coverage of non-English literature, various inconsistencies (e.g. indexation of non-existing journals) in the data (Clermont and Dyckhoff, 2012), and a lack of transparency of the coverage (Wouters and Costas, 2012). Furthermore, López-Cózar et al. (2014) have shown that the citation data in Google Scholar can easily be manipulated by researchers who would want to increase their citations count. According to Wouters and Costas (2012), Google Scholar seems "to be more useful for self-assessment than for systematic impact measurements at several levels of aggregation".

Amongst the other existing tools, there are citation indexes that aim at a comprehensive coverage of specific fields. Some examples are Chemical Abstracts Services, which provides an extensive coverage of chemistry literature, and CAB Abstracts, covering agriculture, environment, veterinary sciences, applied economics, food science and nutrition. Such tools provide a more complete portrait of the scientific production in the fields covered than interdisciplinary databases like WoS and Scopus, and thus may be better suited for field specific research evaluation. Also, as we have shown, most countries and languages are underrepresented in WoS and Scopus, which contributes to the known lack of visibility of research done in some of countries. Many countries have developed national citation indexes in order to address this issue (e.g. Indian Citation Index (ICI), Serbian Citation Index (SCIndeks), Thai-Journal Citation Index (TCI)). These national citation indexes provide a more complete picture of the research done by local scientists, and they also make it possible to assess more accurately the impact of research at the national level.

The advantages that field specific or national citations indexes have over multidisciplinary and international indexes like WoS and Scopus may make them the best tools for certain types of analyses. However, they do not seem to provide a suitable alternative to WoS and Scopus when it comes to performing multidisciplinary and international bibliometric analyses. In other words, when using bibliometric methods for research evaluation, what matters is to understand what each tool has to offer and what its limits are, and to choose the right tool for the task. This study looked at the coverage of WoS and Scopus in order to provide a better view of their coverage characteristics. We have shown that despite the larger coverage of Scopus, the coverage in both databases is unbalanced between countries and languages and that it may introduce some biases when performing comparative analyses. Those are important characteristics that should at least be taken into account when drawing conclusions using these tools for bibliometric analyses, and perhaps more importantly, for research evaluation purposes.

## Acknowledgements

The authors would like to thank Vincent Larivière for his guidance and insights.

## Cited references